\documentclass[aps,prl,reprint,superscriptaddress,footinbib,twocolumn]{revtex4}
\usepackage{dsfont}
\usepackage{amsmath}
\usepackage{graphicx}
\usepackage{multirow}
\usepackage[latin1]{inputenc}
\usepackage{color}
\newcommand{\braketone}[1]{\langle#1\rangle}

\newcommand{\Teff}{T_{\mathrm{eff}}}

\begin{document}


\title{Relaxation and Pre-thermalization in an Isolated Quantum System}



\author{M.~Gring}
\author{M.~Kuhnert}
\author{T.~Langen}
\affiliation{Vienna Center for Quantum Science and Technology, Atominstitut, TU Wien, Stadionallee 2, 1020 Vienna, Austria}

\author{T.~Kitagawa}
\affiliation{Harvard-MIT Center for Ultracold Atoms, Department of Physics, Harvard University, Cambridge, Massachusetts 02138, USA}

\author{B.~Rauer}
\author{M.~Schreitl}
\affiliation{Vienna Center for Quantum Science and Technology, Atominstitut, TU Wien, Stadionallee 2, 1020 Vienna, Austria}

\author{I.~Mazets}
\affiliation{Vienna Center for Quantum Science and Technology, Atominstitut, TU Wien, Stadionallee 2, 1020 Vienna, Austria}
\affiliation{Ioffe Physico-Technical Institute, 194021, St. Petersburg, Russia}

\author{D.~Adu~Smith}
\affiliation{Vienna Center for Quantum Science and Technology, Atominstitut, TU Wien, Stadionallee 2, 1020 Vienna, Austria}

\author{E.~Demler}
\affiliation{Harvard-MIT Center for Ultracold Atoms, Department of Physics, Harvard University, Cambridge, Massachusetts 02138, USA}

\author{J. Schmiedmayer}
\affiliation{Vienna Center for Quantum Science and Technology, Atominstitut, TU Wien, Stadionallee 2, 1020 Vienna, Austria}
\email[]{schmiedmayer@atomchip.org}


\date{\today}

\begin{abstract}
Understanding relaxation processes is an important unsolved problem in many areas of physics. A key challenge in studying such non-equilibrium dynamics is the scarcity of experimental tools for characterizing their complex transient states. We employ measurements of full quantum mechanical probability distributions of matter-wave interference to study the relaxation dynamics of a coherently split one-dimensional Bose gas and obtain unprecedented information about the dynamical states of the system. Following an initial rapid evolution, the full distributions reveal the approach towards a thermal-like steady state characterized by an effective temperature that is independent from the initial equilibrium temperature of the system before the splitting process. We conjecture that this state can be described through a generalized Gibbs ensemble and associate it with pre-thermalization.
\end{abstract}

\maketitle

Despite its fundamental importance, a general understanding of how isolated quantum many-body systems approach (thermal) equilibrium is still elusive. While theoretical concepts such as the quantum ergodic theory or the eigenstate thermalization hypothesis \cite{Polkovnikov2011,Goldstein2010a,Rigol2008} try to infer requirements for a system to be able to undergo relaxation, it is still unclear on what timescale this will happen. Most prominently, in situations where conservation laws or the presence of many constants of motion inhibit efficient relaxation, many-body systems are expected to display a complex behavior. An intriguing phenomenon which has been suggested in this context is \textit{pre-thermalization} \cite{Berges2004}. It predicts the rapid establishment of a quasi-stationary state that differs from the real thermal equilibrium of the system. Full thermalization, i.e. relaxation to the real thermal equilibrium, if present at all, is expected to occur on a much longer time scale.
Pre-thermalized states have been predicted for a large variety of physical systems \cite{Eckstein2009,Moeckel2010,Mathey2010,Barnett2011} and it is conjectured that they can be described by equilibrium statistical mechanics through a generalized Gibbs ensemble\,\cite{Rigol2008,Kollar2011,Polkovnikov2011}. Here we present a direct observation of such a state.

Systems of ultracold atoms provide unique opportunities to experimentally study such non-equilibrium problems because of their almost perfect isolation from the environment. Moreover the timescales for internal relaxation processes (collisions) are easily accessible in experiments.  Consequently, there have recently been various studies about non-equilibrium dynamics in ultracold atom systems \cite{Kinoshita2006,Hofferberth2007a,Sadler2006,Ritter2007,Widera2008,Trotzky2012}.

\begin{figure*}
    \centering
        \includegraphics[width=0.99\textwidth]{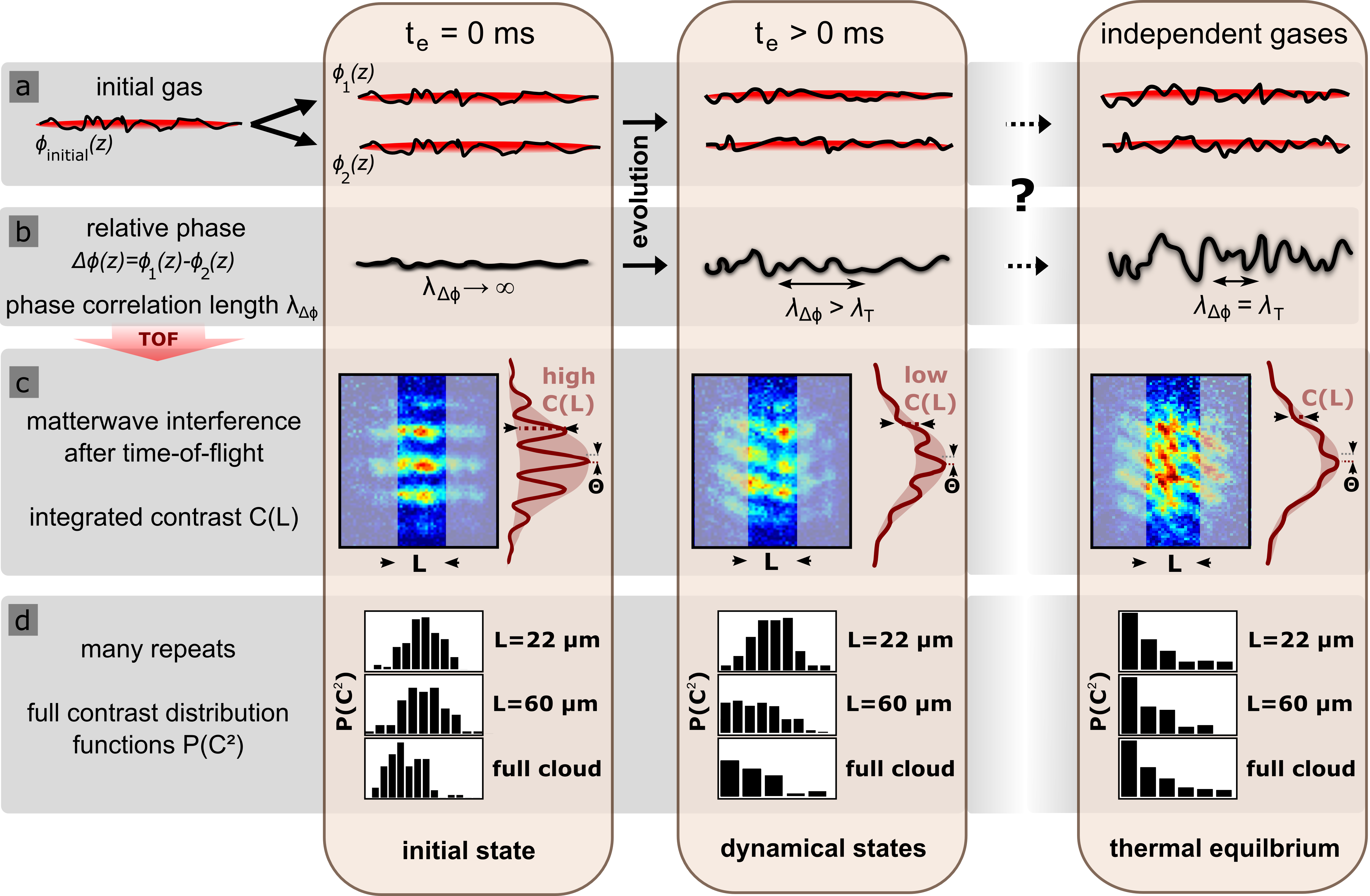}
    \caption{Experimental scheme. \textbf{(a)} An initial phase fluctuating 1d Bose gas is split into two uncoupled gases with almost identical phase distributions $\phi_1(z)$ and $\phi_2(z)$ (represented by the black solid lines) and allowed to evolve for a time $t_{e}$. 
    \textbf{(b)} At $t_{e}=0\,$ms, fluctuations in the local phase difference $\Delta \phi(z)$ between the two gases are very small and the corresponding phase correlation length is very large.  During the evolution these relative phase fluctuations increase and the correlation length decreases. The main question we address in this paper is, if or when this system will reach the corresponding thermal equilibrium of uncorrelated phases as characterized by the initial temperature T, and thermal coherence length $\lambda_T$. In experiment, this situation can be prepared on purpose by splitting a thermal gas and cooling it into two independent gases \cite{Hofferberth2006}. 
    \textbf{(c)} shows typical experimental matter-wave interference patterns obtained by overlapping the two gases in time-of-flight after different evolution times. Differences in the local relative phase lead to a locally displaced interference pattern. Integrated over a length L, the contrast C(L) in these interference patterns is a direct measure of the strength of the relative phase fluctuations. 
    \textbf{(d)} Due to the stochastic nature of the phase distributions, repeated experimental runs yield a characteristic distribution $P(C^2)$ of contrasts, which allows one to distinguish between the initial state, an intermediate pre-thermalized state and the true thermal equilibrium of the system.
    }
    \label{fig:expmeth}
\end{figure*}

One-dimensional (1d) Bose gases are of particular interest because they inherently contain strong fluctuations and dynamics: At finite temperature many longitudinal modes of the system are populated which manifests itself in the rich spatial structure and dynamics in their local phase. This is in stark contrast with three-dimensional condensates, where the existence of long-range order allows to characterize the state with a single, global phase.
In addition a homogeneous 1d Bose gas is a prime example of an integrable quantum system \cite{Lieb1963}.  The near-integrability of experimentally realized trapped 1d Bose gases thus opens up the possibility of studying  dynamics and relaxation close to an integrable point. 

In our experiment (Fig.\,1) we start from a single 1d Bose gas of $^{87}$Rb in the quasi-condensate regime\,\cite{Kheruntsyan2003} prepared in an elongated microtrap on an atom chip\,\cite{AtomChips2011}. We prepare the initial state for our evolution by rapidly and coherently splitting the single 1d gas, producing a system of two uncoupled 1d Bose gases in a double-well potential.  The two trapped 1d gases only differ by the \emph{quantum shot-noise} introduced in the splitting  (Figs.\,1a). They have almost identical longitudinal phase profile, and are therefore strongly correlated in their phase.  In contrast, two independently created quasi-condensates have vastly different and uncorrelated phase profiles (Fig.\,1 \emph{right column}).  The strongly correlated phase of the two gases after splitting reflects the memory that they originally come from a single quasi-condensate. Our experiment studies how this memory about the the initial state evolves, decays in time, 
and if a thermal equilibrium state corresponding to two independent and classically separated quasi-condensates is reached in the evolution.

We probe the evolution of the local phase difference between the two quasi-condensates by matter-wave interference (Fig.\,1c). The system is allowed to evolve in the double-well for some time $t_e$ before the two 1d gases are released from the trap and allowed to interfere in time-of-flight. The local phase difference along the axial length of the system directly translates to a shift of the interference peaks (Fig.\,1c). To probe the strength of the fluctuations in the local phase difference $\Delta \phi(z)$, we integrate the interference pattern longitudinally over a variable length $L$ and extract from the resultant line profile our main experimental observable: the integrated contrast $C(L)$ (Fig.1c). For the initial state, the local phase difference is close to zero everywhere along the quasi-condensates, and thus the integrated interference contrast $C(L)$ is large for all integration lengths $L$. During the course of the evolution, the phase difference varies in the longitudinal direction due to the strong fluctuations inherent in 1d systems, which results in the reduction of $C(L)$ starting with long integration lengths. Thus the measurements of $C(L)$ allow the characterization of the unique dynamics of 1d quasi-condensates.

\begin{figure}
    \centering
        \includegraphics[width=0.49\textwidth]{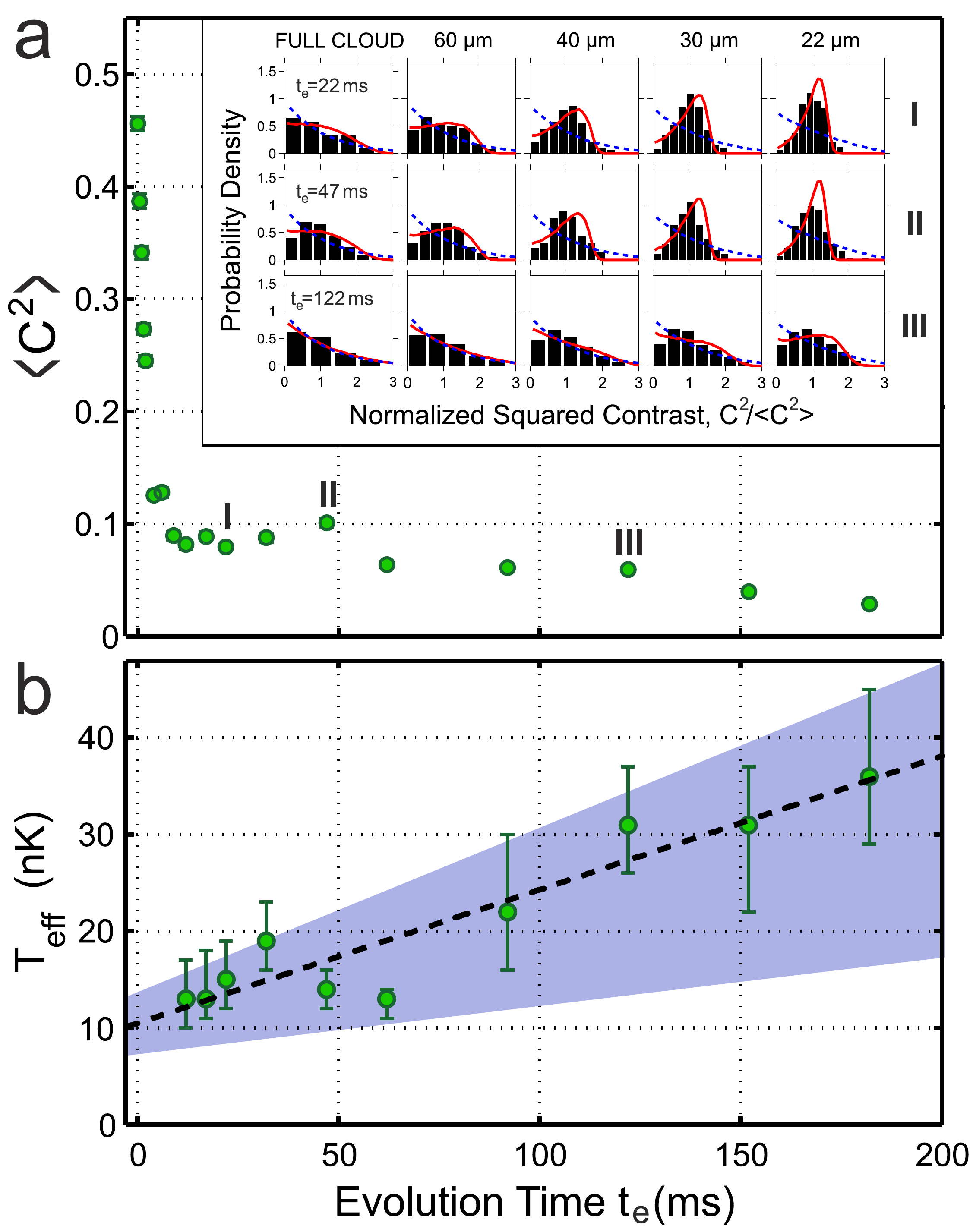}
    \caption{\textbf{(a)} Evolution of the mean squared-contrast $\braketone{C^2}$ for interference patterns integrated over the whole length of the 1d systems. We observe a rapid decay followed by a long slow further evolution.  Error bars are standard errors of the mean. \emph{inset}: Experimental \textit{non-equilibrium} distributions of $C^2/\braketone{C^2}$ at $t_e = 22\,$ms, $47\,$ms, and $122\,$ms respectively (histograms) and a fit of a theoretical \textit{equilibrium} distributions leading to  $T_{\textrm{eff}}=15\pm^4_3\,$nK, $14\pm^2_2\,$nK, and $31\pm^5_6\,$nK respectively (red solid line).  For comparison the calculated \textit{equilibrium} distributions for $T=78\pm10\,$nK (blue dashed line) are added.
    \textbf{(b)}: Evolution of $\Teff$ for the whole data set extracted by fitting equilibrium distributions.  A linear fit indicates an increase of $\Teff$ over time of $0.14\pm0.04\,$nK/ms. The yellow area indicates the measured heating rate of our atom trap of $0.11\pm0.06\,$nK/ms
    }
    \label{fig:Fig2}
\end{figure}

The mean squared contrast $\braketone{C(L)^2}$, similarly to the coherence factor used in \cite{Hofferberth2007a},  is a direct measure of the integrated two-point correlation function of the relative phase between the two halves of the system\,\cite{Polkovnikov2006,Gritsev2006}. 
Integrating over the whole length of the interference pattern we observe (Fig.\,2a) an initial rapid decay in $\braketone{C^2}$ on a time scale of approximately $10\,$ms, after which a quasi-steady state emerges which slowly evolves further on a second, much slower time scale.

The initial rapid decay in Fig.\,2a is analogous to the one observed in the experiment presented in \cite{Hofferberth2007a} which was limited to evolution times $t_e < 12\,$ms by longitudinal dynamics introduced in the splitting. Significant improvements in the experimental techniques (see SOM) allowed us now to reveal the long time behavior.  In the following we will first show that this steady state is not the expected thermal equilibrium and associate it with pre-thermalization \cite{Berges2004}.

To probe the nature of this quasi-steady state we start by employing tools developed to characterize equilibrium systems \cite{Polkovnikov2006,Gritsev2006,Stimming2010,Hofferberth2008} and capture higher-order correlations in the system through the higher moments $\braketone{C^{2n}}$.   For this we extract the full quantum mechanical probability distribution function (FDF) $P(C^2)dC^2$, which gives the probability to observe a value $C^2$ in the interval between $C^2$ and $C^2+dC^2$. The higher moments $\braketone{C^{2n}}$ are directly related to $P(C^2)$ by $\braketone{C^{2n}}=\int{C^{2n} P(C^2) dC^2}$.  Consequently the FDF is a direct measure of \textit{all} even relative phase correlations between the gases and hence determines the state of the system with unprecedented detail \cite{Polkovnikov2006,Gritsev2006}. In particular, high phase coherence between the two halves of the system results in a peaked Gumbel-like distribution, whereas the distribution is exponential in form when the phase coherence is low \cite{Polkovnikov2006,Gritsev2006,Stimming2010,Hofferberth2008}.

Using a statistically large set of data we can map the time evolution of the FDFs for different length scales $L$. For times $>12$\,ms, i.e. directly after the initial rapid evolution shown in Fig.\,2a, we find remarkable agreement of the measured FDFs with theoretical \textit{equilibrium} distributions. We extract an \textit{effective} temperature $T_{\textrm{eff}}$ from a simultaneous fit to the measured data on all length scales probed (see insets in Fig.~\ref{fig:Fig2}a). Surprisingly, immediately after the fast decay at $t_e$ = 12, 17, 22\,ms we find: $T_{\textrm{eff}}=13\pm^4_3$, $13\pm^5_2$, $15\pm^4_3$\,nK respectively, which is more then a factor of five lower than the initial temperature of the un-split system ($T=78\pm10\,$nK). The observed steady-state hence cannot be the true thermal equilibrium state of the system. 
(For a direct comparison of FDFs for the thermal equilibrium distributions in the same double-well system, see SOM.)

In contrast, for evolution times $t_e < 12$\,ms the shapes of the measured FDFs are not consistent with equilibrium theory. The thermal-like appearance of the state is established only during the evolution of the system.

To analyze the subsequent further slow evolution observed in Fig.\,\ref{fig:Fig2}a, we extract the \textit{effective} temperature  for all times after the initial decay. The measured values of $\Teff$ are plotted in Fig.\,\ref{fig:Fig2}b. We find an increase of $\Teff$ over time of $0.14\pm0.04\,$nK/ms. This is, however, consistent with the measured heating rate of our atom trap of $0.11\pm0.06\,$nK/ms  which we characterized independently using equilibrium quasi-condensates (see SOM). This indicates that either no thermalization is present in this nearly integrable system, or, if it is present, that it is a very slow process.

\begin{figure}
    \centering
        \includegraphics[width=0.49\textwidth]{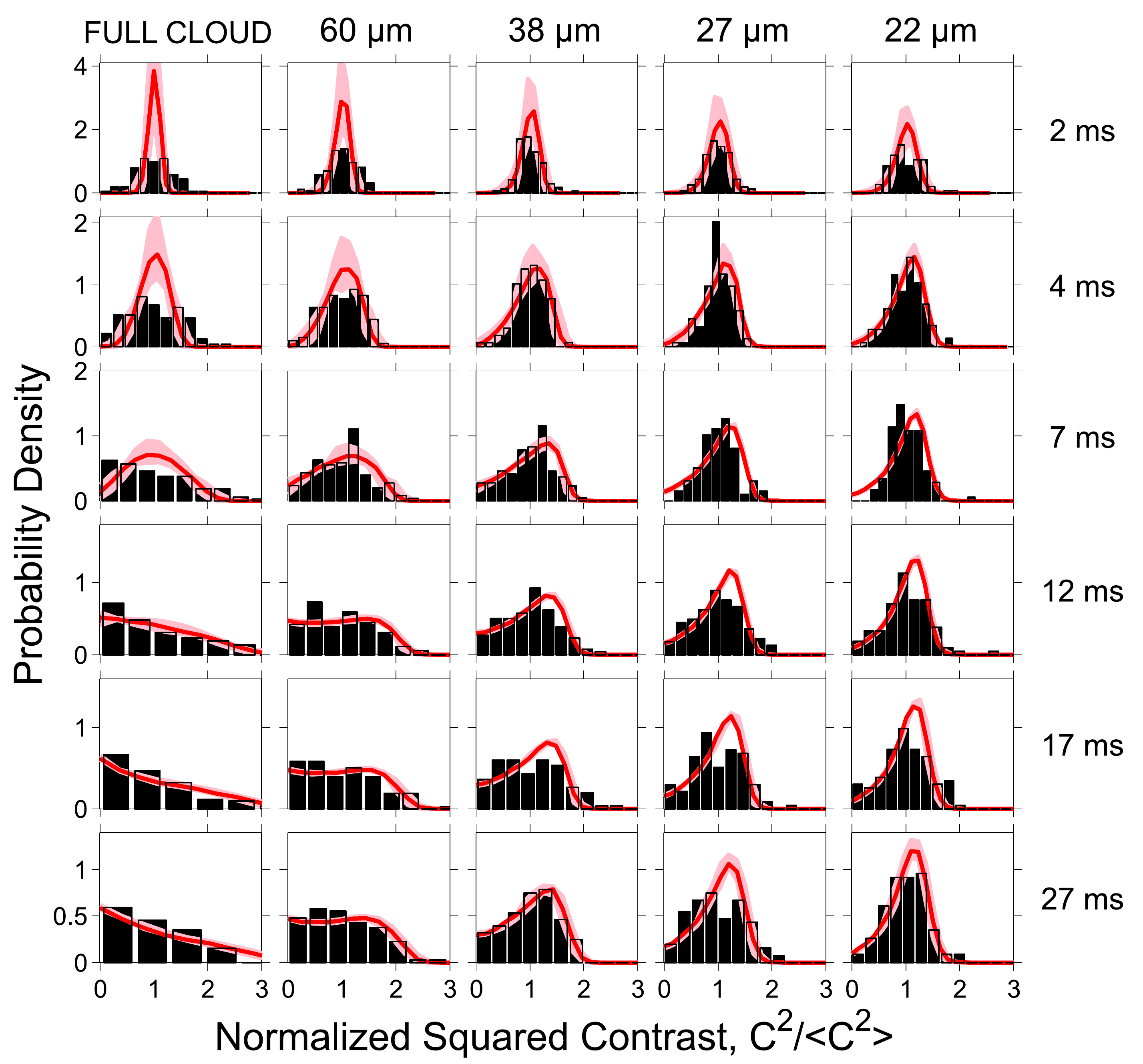}
    \caption{   
     The evolution of the squared-contrast distribution $P(C^2)$ for different integration lengths $L$. Experimental data is plotted using histograms and the theoretical simulations as (red) curves. For integration over the full cloud length, the distribution rapidly becoming exponential with increasing evolution time $t_e$.  For the shortest integration length, the distribution preserves a non-zero peak showing a persisting memory of the correlations of the initial state \cite{Olshanii2012}. The light (pink) shaded areas denote the errors of the experimentally measured theory input parameters.  
    }
    \label{fig:c2distribution}
\end{figure}

To describe the fast evolution from the splitting to the emergence of the quasi-steady state, we employ a fully integrable theory based on a Tomonaga-Luttinger liquid formalism \cite{Kitagawa2010, Kitagawa2011} (for details see SOM). The evolution of the \emph{local phase difference}  between the two halves of the system $\Delta\phi(z)$ is thereby described by a set of uncoupled collective modes with momentum $k$, i.e. sound waves, which modulate the relative density and phase at a wave-length $\lambda=2\pi/k$ and with an amplitude given by the population of the mode.  A sudden splitting creates an equipartition of energy between all the $k$-modes, which initially are all in phase (SOM). The rapid evolution of the system seen over the first $\sim 10\,$ms is then the \emph{dephasing} of these $k$-modes. The FDFs calculated by this integrable theory \cite{Kitagawa2010, Kitagawa2011}, using input parameters independently extracted from the experiment, show remarkable agreement without any free parameter (Fig.\,\ref{fig:c2distribution}).

The model also predicts a steady state, to which the integrable system will relax: The {dephasing}, along with the equipartition of energy between the $k$-modes introduced by the fast splitting, results in the FDFs of the quasi-steady state being indistinguishable from those of a system in thermal equilibrium at some effective temperature $\Teff$, which is determined by the energy given to the relative degrees of freedom by the quantum shot noise introduced in the splitting.  The full calculation gives \cite{Kitagawa2011} 
\begin{equation}
	k_B \Teff = g \rho/2 \,,
	\label{Eq:Teff}
\end{equation}	
where $g=2\hbar\omega_\perp a_s$ is the 1d interaction strength for particles with scattering length $a_s$ trapped in a tube with transversal confinement $\omega_\perp$ and $\rho$ is the 1d density of each half of the system.  For the parameters used in the data presented in Fig.\,\ref{fig:c2distribution} the model predicts $\Teff=11\pm3$\,nK, in very good agreement with our observations of  $\Teff=14\pm4\,$nK, $17\pm5\,$nK, and $14\pm4\,$nK for the evolution times of $12\,$ms, $17\,$ms, and $27\,$ms, respectively.

\begin{figure}
    \centering
        \includegraphics[width=0.49\textwidth]{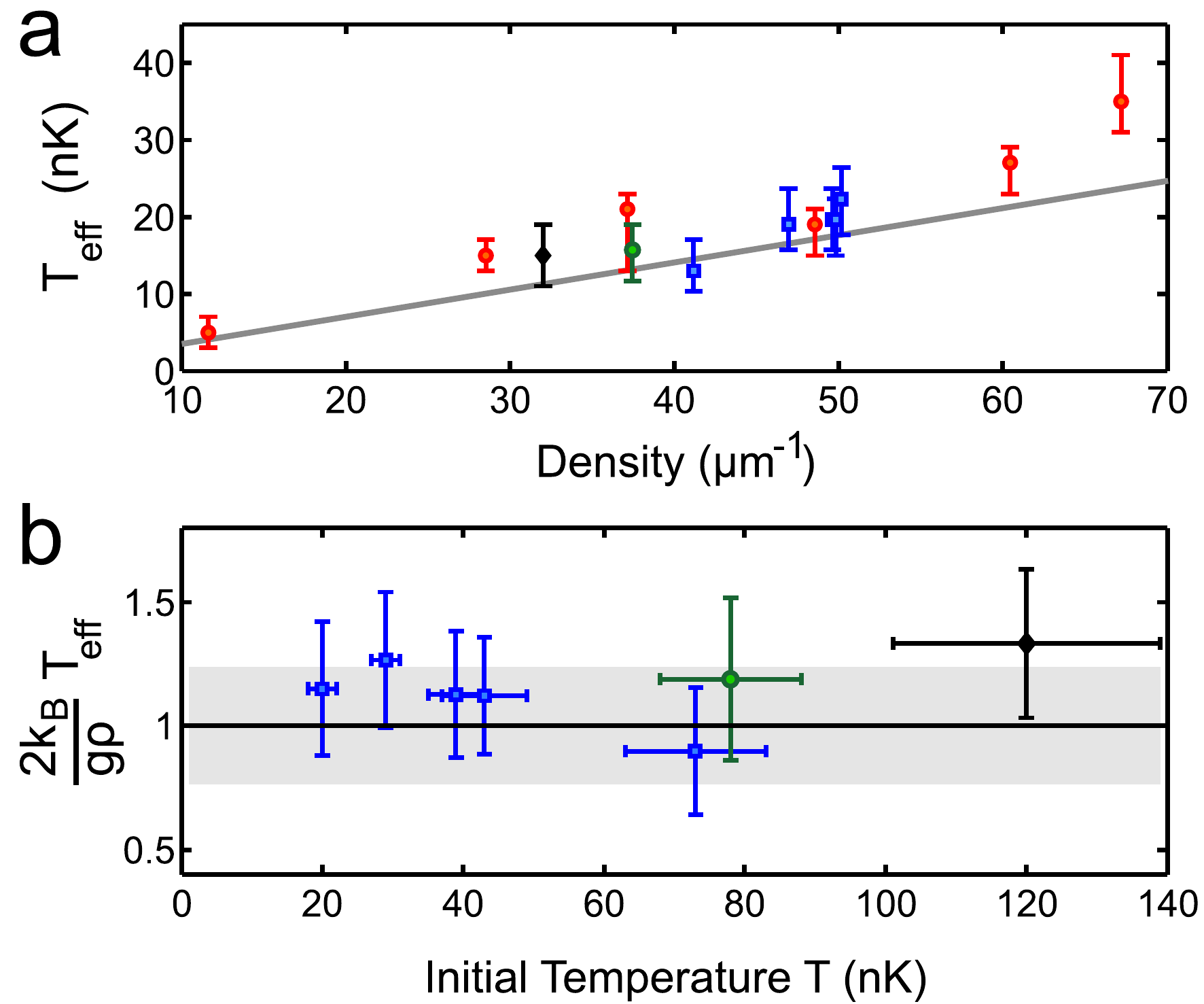}
    \caption{
        \textbf{a)}~Dependence of $\Teff$ on $\rho$ and \textbf{b)}~Independence of $\Teff$ from the initial temperature $T$ of the system before splitting, corrected for the scaling of $\Teff$ with density. The colors encode different datasets.  The (black) solid line corresponds to the theoretical prediction $k_B \Teff= g \rho/2$. 
     The \emph{black} (\emph{green}) data point in \textbf{a} and \textbf{b} correspond to the dataset presented in Fig.\,3, (Fig.\,2) respectively.}
    \label{fig:T_eff dependence}
\end{figure}

Moreover our integrable model predicts (Eq. \ref{Eq:Teff}) that the effective temperature should be linearly dependent on the initial 1d density, and independent of the initial temperature. Both these predictions are confirmed by extending the experiments over a wide range of initial conditions (Fig. \ref{fig:T_eff dependence}). 

The apparent systematic offset of the experimentally derived $\Teff$ and the theoretical prediction in both Figs. \ref{fig:T_eff dependence}a and \ref{fig:T_eff dependence}b can be attributed to imperfections in the experimental splitting process, which in the model is assumed to be instantaneous. This finite-time splitting is also the reason that the agreement between the experiment and theory in Fig. \ref{fig:c2distribution} is less good for very early times.
   
Nevertheless the first milliseconds of the observed dynamics are well-captured by the \emph{integrable} Luttinger liquid theory. The large number of conserved quantities in this integrable system prevents thermalization. Our experimental realization of a 1d system is however not completely integrable and will eventually thermalize. 

Dynamics beyond the harmonic Luttinger Liquid model is required to couple symmetric and anti-symmetric modes and give rise to thermalization \cite{Burkov2007}. A mechanism that is expected to come into play is interactions of particles that go beyond two-body collisions, like three-body processes connected with higher radial trapping states~\cite{Mazets2008,Mazets2010,Tan2010}. In our present experiment, these processes that can lead to full thermalization are much slower than the de-phasing of the collective modes, and thus allow the clear observation of the dynamics dominated by the close-by integrable system. 
It is of great interest to investigate the physics of thermalization in the future and study, for example, how far away from integrability one has to go to see full thermalization and probe its time-scale.

In view of our present analysis, the observed decay of the coherence factor in the experiment of Hofferberth et al.\,\cite{Hofferberth2007a} has to be reinterpreted.  In agreement with our present experiment it shows the same fast 'integrable' de-phasing of relative modes in the spit 1d system\,\cite{Kitagawa2010,Kitagawa2011,Bistritzer2007}, and not full decoherence and thermalization as originally interpreted by comparison with the theoretical description of Burkov et al.\,\cite{Burkov2007} (for more details see SOM)
We point out that for the present experiment, even independent of our theoretical model, the observed independence of $\Teff$ from the initial temperature provides direct experimental evidence that we do not observe thermalization.

In summary, the quasi-steady state found in our experiments provides the first direct observation of pre-thermalization, as predicted to appear in non-equilibrium systems close to an integrable point~\cite{Kollar2011}. We would like to point out that we observe pre-thermalization in the relative degrees of freedom. The initial thermal energy is still stored in the common mode fluctuations of the two halves of the system which is are not probed by the interference pattern.

The effective temperature being significantly different from the kinetic temperature supports the prediction that such a state requires a description by a generalized Gibbs ensemble\,\cite{Rigol2008,Kollar2011,Polkovnikov2011}. 

We note that the thermal-like appearance of the pre-thermalized state is not necessarily generic but a special property of our system and is due to the fast splitting process which puts equal energy into all (relative) collective modes of the system. 

Our experiment also directly shows that the two separated many-body systems retain memory of their initial state for a time much longer than the randomization of the global phase would suggest, and that genuine de-coherence which would erase the memory did not yet occur, i.e. the two 1d systems did not yet emerge as two classically separated entities.

The timescale over which this pre-thermalized state persists remains an open question.  It is directly related to the open problems of how two quantum-mechanically entangled objects reach classicality, the properties of the hypothetical quantum KAM theorem \,\cite{Polkovnikov2011}, and the very nature of thermalization itself.

 We thank Ch. v. Hagen and M. G\"obel for early work on the experimental apparatus and the Vienna group for invaluable discussions and assistance.  The atom chip was fabricated at the ZMNS, TU Wien by W. Schrenk and M. Trinker. The experiments were supported by the Austrian FWF through \textit{M1040-N16}, the Doctoral Programme CoQuS (\textit{W1210}) and the the Wittgenstein Prize, the EU through the integrating project AQUTE and \textit{P22590}, Siemens Austria, and The City of Vienna.  TK and ED thank the Army Research Office for funding from the DARPA OLE, program, Harvard-MIT CUA, NSF Grant No. DMR-07-05472, AFOSR Quantum Simulation MURI, and the ARO-MURI on Atomtronics. 

\vspace{-5mm}
\bibliography{NEMBD} 

\end{document}